\documentclass{jfm}
\usepackage{graphicx}
\usepackage{color}
\usepackage{epstopdf, epsfig}
\usepackage{amsmath}

\newcommand{\bs}{\boldsymbol}

\newcommand{\change}{\color{red}}

\shorttitle{No net motion for oscillating near-spheres at low {Reynolds} numbers}
\shortauthor{K.Lippera, O. Dauchot, S. Michelin, M. Benzaquen}

\title{No net motion for oscillating near-spheres at low {Reynolds} numbers}

\author{K. Lippera\aff{1},
  O. Dauchot\aff{2},
  S. Michelin\aff{1}
    \and M. Benzaquen\aff{1} \corresp{\email{michael.benzaquen@polytechnique.edu}}}

\affiliation{\aff{1}LadHyX,  UMR CNRS 7646, Ecole polytechnique,  91128 Palaiseau, France
\aff{2}EC2M, UMR CNRS 7083 Gulliver, ESPCI ParisTech, 10 rue Vauquelin, 75005 Paris, France}

\begin{document}

\maketitle

\begin{abstract}
We investigate the flow around an oscillating nearly-spherical particle at low, yet non-vanishing, Reynolds numbers, and the potential resulting locomotion. We analytically demonstrate that no net motion can arise up to order one in Re and order one in the asphericity parameter, regardless of the particle's shape. Therefore, geometry-induced acoustic streaming propulsion, if any, must arise at higher order.
\end{abstract}

\section{Introduction}

Solid bodies forced to oscillate in a fluid may, as a result, undergo a net motion, provided their shape breaks an appropriate symmetry. Yet, in the absence of inertia (i.e. when the Reynolds number $\mbox{Re}$ is strictly zero), no net motion can arise from time-reciprocal actuation due to the linearity of Stokes' equations~\citep{purcell1977}. Above a critical $\mbox{Re}_c=O(1)$, a symmetric rigid body can achieve unidirectional locomotion as a result of symmetry-breaking 
instability resulting from the nonlinear inertial contribution to the Navier-Stokes equations~\citep{alben2005coherent}. The purpose of the present work is to analyze the emergence of self-propulsion at small but finite $\mbox{Re}$ (i.e. the effect of inertia is weak but non-negligible) for oscillating asymmetric particles. Indeed one could expect that  asymmetric flows, resulting from asymmetric boundary conditions, shall push the particle, thereby inducing nonzero average motion~\citep{nadal2014asymmetric}.

Artificial microswimmers have received much recent attention, thanks to their potential application to drug delivery or water treatment \citep{sundararajan2008catalytic,martinez2015magnetic,tiwari2008application}, or their fundamental interest to study active matter~\citep[see e.g.][]{Palacci:2013eua,Buttinoni:2013dea,Bechinger:2016cf}. Among the many possible routes to self-propulsion, swimming in self-generated physico-chemical gradients i.e. autophoresis~\citep{moran2017}, as well as bubble-generating~\citep{wang2014,li2016} or magnetically-actuated microswimmers~\citep{dreyfus2005microscopic} have received a particular attention. In these examples, a front-back asymmetry in the design of the system is necessary. Yet, symmetry-breaking and self-propulsion can also be achieved by exploiting an instability \citep{bricard2013emergence,Michelin:2013aa,izri2014self} or flexibility~\citep{wiggins1998flexive}.

Recently, passive rigid particles levitating in the nodal planes of an acoustic stationary wave have been observed to self-propel in a plane orthogonal to their direction of excitation~\citep{wang2012autonomous}. To explain such findings,  \citet{nadal2014asymmetric} proposed an  acoustic streaming mechanism, suggesting that near-spherical particles with asphericity parameter $\epsilon$ can achieve a net $O(\epsilon\mbox{Re})$ propulsion, in the low frequency limit. Several studies have since stood upon {the results of Nadal \& Lauga  to account for their observations}  \citep[see e.g.][]{sabrina2018shape,ahmed2016density,soto2016acoustically}. 

In contrast, we here demonstrate analytically that no net motion can arise at $O(\epsilon\mbox{Re})$ from a time-reciprocal oscillation and that  self-propulsion, if any,  must arise at higher order. In Section \ref{goveq}, the governing equations for an oscillating particle are presented. In Section \ref{nsp}, we introduce the particle geometry and the Taylor expansions of the velocity fields in Re and $\epsilon$. In Sections \ref{zeroRe} and \ref{1Re}, we compute the net motion of the particle at the first two orders in Re. In Section \ref{ccl}, we discuss our results and conclude.

\section{Governing equations}\label{goveq}
We consider here a rigid {and homogeneous} particle of  typical size $R$ oscillating with frequency $\omega$ and amplitude $a$ in an incompressible and Newtonian fluid of kinematic viscosity $\nu$. Using $R$, $a\omega$ and $1/\omega$ respectively as reference length, velocity and time scales, the dimensionless Navier-Stokes and continuity equations read \citep{zhang1998oscillatory}: 
\begin{equation}
\label{navierstokes}
\lambda^2\partial_t \bs u+\mbox{Re} \bs \nabla \bs u \cdot \bs u=\bs \nabla\cdot \bs \sigma \ ,\qquad 
\bs\nabla\cdot\bs u=0 \ ,
\end{equation}
with $\bs \sigma = -p \bs I + (\bs \nabla \bs u + \bs {\nabla^\top} \bs u)$, the dimensionless stress tensor. The Reynolds number and reduced frequency are respectively defined as $\mbox{Re}=a\omega R / \nu$ and $\lambda^2=(R/\delta)^2$ with $\delta=\sqrt{\nu/\omega}$ the viscous penetration depth. More precisely, a translational oscillation is imposed to the particle along the $\bs e_x$ direction, $\tilde{\bs U}=e^{it}\bs e_x$, and the particle is free to move along the other directions, and is thus force-free along the $yz$ plane and torque-free about any axis.  The longitudinal and angular velocities of the particle resulting from its imposed oscillation are $\bs U=U_y \bs e_y + U_z \bs e_z$ and $\bs \Omega=\Omega_x \bs e_x+\Omega_y \bs e_y+\Omega_z \bs e_z$. In the frame of reference of the laboratory, the boundary conditions read: 
\begin{equation}
\label{BCfo1}
\bs u|_S=\tilde{\bs U}+\bs U+\bs \Omega\times \bs r\ ,\qquad
\bs u|_{\bs r \rightarrow \infty}=\bs 0 \ .
\end{equation}
In order to determine $\bs U$ and $\bs \Omega$ following an approach analogous to that of Lorentz' reciprocal theorem \citep{happel1965low}, the auxiliary problem of a particle of the same \emph{instantaneous} geometry in a steady Stokes flow is considered:
\begin{equation}
\label{unstokes}
\bs \nabla\cdot \widehat{\bs \sigma} = \bs 0\ ,\qquad 
\bs\nabla\cdot\widehat{\bs u}=0 \ ,
\end{equation}
with boundary conditions,
\begin{equation}
\label{BCzo1}
\widehat{\bs u}|_S=\widehat{\bs U}+\widehat{\bs \Omega}\times \bs r\ ,\qquad
\widehat{\bs u}|_{\bs r \rightarrow \infty}=\bs 0 \ .
\end{equation}
Let us stress that the particle is rigid so that by {instantaneous} geometry one should understand that the surface boundary of the auxiliary problem matches that of its real counterpart at each time.
Using \eqref{navierstokes} and  \eqref{unstokes}, one obtains:
\begin{eqnarray}
\int_V\Big[\widehat{\bs u}\cdot(\bs\nabla\cdot\bs\sigma)-\bs u\cdot(\bs\nabla\cdot\widehat{\bs\sigma})\Big]\mathrm{d}V=\mbox{Re}\int_V \widehat{\bs u} \cdot  \bs \nabla \bs u \cdot \bs u\, \mathrm{d}V + \lambda^2 \int_V \widehat{\bs u}\cdot \partial_t \bs u \,\mathrm{d}V \ . \quad \label{vw}
\end{eqnarray}
 Using the divergence theorem together with the continuity equations, \eqref{vw} reduces to:
\begin{eqnarray}
\int_{S_{\infty}-S}\left(\widehat{\bs u}\cdot\bs\sigma-\bs u\cdot\widehat{\bs\sigma}\right)\cdot \bs n \,\mathrm{d}S=\mbox{Re}\int_V \widehat{\bs u} \cdot  \bs \nabla \bs u \cdot \bs u\,\mathrm{d}V + \lambda^2 \int_V \widehat{\bs u}\cdot \partial_t \bs u \,\mathrm{d}V \ .\label{vwdt}
\end{eqnarray}
Because $\bs u, \widehat{\bs u} \sim 1/r$ and $\bs \sigma,\widehat{\bs \sigma} \sim 1/r^2$  when $r\to \infty$ \citep[see e.g.][]{happel1965low}, the surface integral at infinity in \eqref{vwdt} vanishes. The boundary conditions \eqref{BCfo1} and \eqref{BCzo1} then yield:
\begin{equation}
(\tilde{\bs U}+\bs U)\cdot \widehat{\bs F} + \bs \Omega \cdot \widehat{\bs L}-\widehat{\bs U}\cdot \bs F-\widehat{\bs\Omega}\cdot \bs L  =\mbox{Re}\int_V \widehat{\bs u} \cdot   \bs \nabla\bs u \cdot \bs u\,\mathrm{d}V + \lambda^2 \int_V \widehat{\bs u}\cdot \partial_t \bs u \,\mathrm{d}V\ ,\quad
\label{pvw}
\end{equation}
with $\bs F=\int_S\bs \sigma \cdot \bs n \mathrm{d}S$ and $\bs L=\int_S(\bs r\times\bs \sigma) \cdot \bs n  \mathrm{d}S$ (resp. $\hat{\bs F}$ and $\hat{\bs L}$), the force and torque in the real (resp. auxiliary) problem. For the real problem, $\bs F$ and $\bs L$ derive from Newton's laws:
\begin{equation}
\label{NL}
\bs F=\overline{\rho}\partial_t{\bs U}\ ,\qquad
\bs L=\partial_t(\bs J \cdot \bs\Omega)\ ,
\end{equation}
with $\bar{\rho}$ the particle-to-fluid density ratio and $\bs J$ the particle's inertia tensor. For the auxiliary problem, $\hat{\bs F}$ and $\hat{\bs L}$ are linearly related to $\hat{\bs U}$ and $\hat{\bs \Omega}$ through the possibly non-diagonal resistance matrix \citep{kim2013microhydrodynamics}.
In order to compute the particle motion $(\bs U,\bs \Omega)$, we shall consider in \eqref{pvw} either (i) an auxiliary steady propulsion $(\hat{\bs U},\bs 0)$ with $\hat{\bs U} \parallel \bs U$ to determine $\bs U$, or (ii) an auxiliary steady rotation $(\bs 0,\hat{\bs \Omega})$ with $\hat{\bs \Omega}\parallel \bs \Omega$ to determine $\bs \Omega$. Note that finding the  contribution at $O(\mbox{Re}^n)$ of the first term on the right-hand side of \eqref{pvw} relies on the knowledge of the  velocity field $\bs u$ at $O(\mbox{Re}^{n-1})$ only, hence the possibility of a recursive calculation order by order in $\mbox{Re}$. Conversely, computing the second term shall rely on peculiar symmetry and time-average considerations to be made explicit below. 
Note that for a homogeneous particle, the above  formulation also applies to the motion of a particle exposed to a uniform oscillating flow $- \tilde{\bs U}$, once inertial forces are accounted for as a modified pressure.

\section{Nearly spherical particles in low Reynolds flows}\label{nsp}
{We consider a nearly-spherical particle of volume $V$ and center-of-mass $O$. By choosing $R=(3V/4\pi)^{1/3}$ and taking $O$ as the origin of the system of axes, one can define the particle's geometry through} $r=1+\epsilon f(\bs n )$ with $\epsilon\ll1$. {By construction $f$ satisfies: }
\begin{equation}
\label{volume}
\int_{S}f(\bs n)\mathrm{d}S=0\ ,\qquad 
\int_{S}f(\bs n) \bs n \mathrm{d}S=\bs 0\ .
\end{equation}
The governing equations are first linearised with respect to $\mbox{Re}\ll 1$, e.g. defining $\bs u=\bs u_0+\mbox{Re} \bs u_1 +O(\mbox{Re}^2)$, and each order is further expanded as a regular perturbation problem  in $\epsilon\ll 1$, e.g. $\bs u_k=\bs u_k^{0}+\epsilon \bs u_k^{\epsilon}+O(\epsilon^2)$ with $k=0,1$~\footnote{Note that $\epsilon$ must remain small compared to all other dimensionless length scales, i.e. $\epsilon\ll 1$ (particle radius) and $\epsilon\ll 1/\lambda$ (viscous boundary layer thickness).}.
In the following sections, we shall consider the problems at $O(\mbox{Re}^k\epsilon^\ell)$, and successively look into the two leading orders in Re.

\section{Zeroth-order in $\mbox{Re}$}\label{zeroRe}

At  leading order $O(\mbox{Re}^0)$, \eqref{navierstokes} and \eqref{BCfo1} become:
\begin{align}
\label{Stokes}
\lambda^2\partial_t \bs u_0=-\bs \nabla p_0+\bs \nabla^2\bs u_0\ ,&\qquad 
\bs\nabla\cdot\bs u_0=0 \ ,\\
\label{bc0}
\bs u_0|_S= \tilde{\bs U} + \bs U_0+\bs \Omega_0\times \bs r\ ,&\qquad 
\bs u_0|_{\bs r \rightarrow \infty}= \bs 0 \ ,
\end{align}
and \eqref{pvw} reduces to:
\begin{eqnarray}
(\tilde{\bs U}+\bs U_0)\cdot \widehat{\bs F} + \bs \Omega_0 \cdot \widehat{\bs L} -\widehat{\bs U}\cdot \bs F_0-\widehat{\bs\Omega}\cdot \bs L_0 =\lambda^2 \int_V \widehat{\bs u}\cdot \partial_t \bs u_0 \,\mathrm{d}V\ ,
\label{Lorentz_0}
\end{eqnarray}
and this result is expanded as a linear perturbation in $\epsilon$ below.

\subsection{{Perfect sphere -- }$O({\normalfont \mbox{Re}}^0\epsilon^0)$}

While it is quite clear that no net motion can arise at {$O(\epsilon^0\mbox{Re}^0)$} (i.e. unsteady Stokes flow around a spherical particle), we briefly rederive this result to provide the reader with the general methodology. 
At leading order $O(\epsilon^0)$, \eqref{Lorentz_0} becomes:
\begin{eqnarray}
(\tilde{\bs U}+\bs U_0^0)\cdot \widehat{\bs F}^0 + \bs \Omega_0^0 \cdot \widehat{\bs L}^0-\widehat{\bs U}\cdot \bs F_0^0-\widehat{\bs\Omega}\cdot \bs L_0^0 =\lambda^2 \int_{V_0} \widehat{\bs u}^{0}\cdot \partial_t \bs u_0^0 \,\mathrm{d}V\ ,
\label{Lorentz00}
\end{eqnarray}
where $V_0$ denotes the volume of fluid outside the reference unit sphere. First, recalling $\widehat{\bs U}\parallel\bs U^0_0$ provides $\widehat{\bs U}\cdot\tilde{\bs U}=0$. Second,  the velocity field $\widehat{\bs u}^{0}$ (resp. $\bs u_0^0$) is linear  with respect to  $\widehat{\bs U}$ (resp. $\tilde{\bs U}$), and  axisymmetric about the axis holding the vector $\widehat{\bs U}$ (resp. $\tilde{\bs U}$) and passing through the centre of mass of the particle. As a result, using the expression of $\widehat{\bs u}^{0}$ and ${\bs u}_0^{0}$ (Appendix~\ref{apA}) shows that the RHS of \eqref{Lorentz00} 
 does not include any contribution from the forcing $\tilde{\bs U}$.  There is therefore no net motion at this order, i.e. $\bs U_0^0=0$. A similar reasoning shows that $\bs\Omega_0^0=0$ as well. This last result imposes the rotation velocity of the particle to be at least first order (either in $\epsilon$ or $\mbox{Re}$). The forcing and induced rotation act therefore on two separate time scales. As a consequence, at leading order, the geometry of the particle, $f$, can be considered constant over the $O(1)$ period of the forcing (fast time scale).

\subsection{{Near-sphere correction -- }$O({\normalfont \mbox{Re}}^0\epsilon^1)$}

At  $O(\epsilon^1)$ 
\eqref{Lorentz_0} becomes: 
\begin{align}
\bs U_0^{\epsilon}\cdot \widehat{\bs F}^0+\bs \Omega_{0}^{\epsilon}\cdot\widehat{\bs L}^0&+\tilde{\bs U}\cdot \widehat{\bs F}^{\epsilon}-\widehat{\bs U}\cdot \bs F_0^{\epsilon}-\widehat{\bs \Omega}\cdot \bs L_0^{\epsilon}\nonumber\\
 &=\lambda^2 \int_{V_0} \left(\widehat{\bs u}^{\epsilon}\cdot \partial_t \bs u_0^0 + \widehat{\bs u}^{0}\cdot \partial_t \bs u_0^{\epsilon}\right) \mathrm{d}V -\lambda^2\int_{S_0}f \widehat{\bs u}^0\cdot \partial_t\bs u_0^0 \,\mathrm{d}S\ ,
\label{Lorentz01}
\end{align}
where the surface integral {is the $O(\epsilon)$ contribution from the difference of the volume integrals on $V$ and $V_0$~\citep[e.g][]{ zhang1998oscillatory}}. The analysis of \citet{zhang1998oscillatory}  shows that the rotation of a torque-free homogeneous near-sphere resulting from an $O(1)$ imposed translation is $O(\epsilon^2)$ and thus $\bs \Omega_0^{\epsilon}=\bs 0$. {Consequently the torque $\bs L_0^{\epsilon}$ linked to $\bs \Omega_0^{\epsilon}$ through Newton's law \eqref{NL} vanishes as well.} Using  \eqref{BCzo1} and \eqref{volume}, the last term in \eqref{Lorentz01} vanishes exactly:
\begin{equation}
\int_{S_0}f \widehat{\bs u}^0\cdot \partial_t\bs u_0^0\, \mathrm{d}S=(\dot{\tilde{\bs U}}\times \widehat{\bs \Omega})\cdot \int_{S_0}f \bs n \,\mathrm{d}S =0\ .
\end{equation}Since we are interested in the net motion of the particle, we take the time-average over the fast time scale (forcing period) of \eqref{Lorentz01}. The  $\langle \mbox{RHS}\rangle_t$ can be shown to vanish because $\bs u_0^{\epsilon}$ and $\bs u_0^{0}$ are periodic in time, and the integration domains are  time-independent. Therefore:
\begin{eqnarray}
\langle\bs U_0^{\epsilon}\rangle_t\cdot \widehat{\bs F}^0-\widehat{\bs U}\cdot \langle\bs F_0^{\epsilon}\rangle_t=0\ .
\label{unsteadyU2}
\end{eqnarray}
Equation \eqref{unsteadyU2} is linear with no net contribution of the forcing $\tilde{\bs U}$: no net motion can occur at  $O({\mbox{Re}}^0\epsilon^1)$, $\langle{\bs U_0^\epsilon}\rangle_t=\bs 0$.

\section{First-order in $\mbox{Re}$}\label{1Re}
At  $O(\mbox{Re}^1)$, \eqref{navierstokes} and \eqref{BCfo1} become:
\begin{align}
\label{Stokes2}
\lambda^2\partial_t \bs u_1+\bs \nabla \bs u_0\cdot \bs u_0=-\bs \nabla p_1+\bs \nabla^2\bs u_1\ ,&\qquad 
\bs\nabla\cdot\bs u_1=0 \ ,\\
\label{bc02}
\bs u_1|_S= \bs U_1+\bs \Omega_1\times \bs r\ ,&\qquad 
\bs u_1|_{\bs r \rightarrow \infty}= \bs 0 \ ,
\end{align}
and \eqref{pvw} reduces to:
\begin{eqnarray}
\bs U_1\cdot \widehat{\bs F} + \bs \Omega_1 \cdot \widehat{\bs L}-\widehat{\bs U}\cdot \bs F_1-\widehat{\bs\Omega}\cdot \bs L_1  =\int_V \widehat{\bs u} \cdot \bs \nabla \bs u_0 \cdot \bs u_0\, \mathrm{d}V + \lambda^2 \int_V \widehat{\bs u}\cdot \partial_t \bs u_1 \,\mathrm{d}V\ .
\label{firstorder}
\end{eqnarray}
Note that here, in addition to the unsteady forcing, the non-linear convective term acts as a source term in \eqref{firstorder}. Because it is quadratic in velocity, one might expect that its average in time is nonzero, which could in turn yield net particle motion. 
\subsection{{Perfect sphere -- }$O({\normalfont \mbox{Re}}^1\epsilon^0)$}\label{order10}
At leading order $O(\epsilon^0)$ \eqref{firstorder} becomes: 
\begin{equation}
\bs U_1^0\cdot \widehat{\bs F}^0 + \bs \Omega_1^0 \cdot \widehat{\bs L}^0-\widehat{\bs U}\cdot \bs F_1^0-\widehat{\bs\Omega}\cdot \bs L_1^0 =\int_{V_0} \widehat{\bs u}^0 \cdot  \bs \nabla \bs u_0^0\cdot \bs u_0^0\,\mathrm{d}V + \lambda^2 \int_{V_0} \widehat{\bs u}^0\cdot \partial_t \bs u_1^0 \,\mathrm{d}V\ .\quad
\label{Lorentz10}
\end{equation}
The symmetry properties of $\widehat{\bs u}^{0}$ and $\bs u_0^0$ (Appendix~\ref{apA}) impose that the first term on the RHS of \eqref{Lorentz10} vanishes. The second term on the RHS vanishes as well because it is the integral of the scalar product between two axisymmetric fields about  orthogonal principal directions. Therefore \eqref{Lorentz10} becomes:
\begin{eqnarray}
\label{U01O01}
\bs U_1^0\cdot \widehat{\bs F}^0 + \bs \Omega_1^0 \cdot \widehat{\bs L}^0 -\widehat{\bs U}\cdot \bs F_1^0-\widehat{\bs\Omega}\cdot \bs L_1^0 =\bs 0\ ,
\end{eqnarray}
implying that, very  much like for  $O({\mbox{Re}}^0\epsilon^0)$, $\bs U_1^{0}=\bs 0$ and $\bs \Omega_1^{0}=\bs 0$.
\subsection{{Near-sphere correction -- }$O({\normalfont \mbox{Re}}^1\epsilon^1)$}
At  $O(\epsilon^1)$, and using \eqref{NL} together with \eqref{U01O01}, \eqref{firstorder} becomes: 
 \begin{eqnarray}
& &\bs U_1^{\epsilon}\cdot\widehat{\bs F}^0+\bs \Omega_1^{\epsilon}\cdot\widehat{\bs L}^0 - \widehat{\bs U}\cdot\bs F_1^{\epsilon}  -     \widehat{\bs \Omega}  \cdot \bs L_1^\epsilon =-\int_{S_0}f \widehat{\bs u}^{0} \cdot  \bs \nabla \bs u_0^0\cdot \bs u_0^0\, \mathrm{d}S\nonumber\\
& &\quad\quad+\int_{V_0} \left(\widehat{\bs u}^{\epsilon} \cdot  \bs \nabla \bs u_0^0\cdot \bs u_0^0+\widehat{\bs u}^{0}\cdot[\bs \nabla \bs u_0^{0}\cdot \bs u_0^{\epsilon} +\bs \nabla \bs u_0^{\epsilon}\cdot \bs u_0^{0}]\right)\mathrm{d}V\nonumber\\
& &\quad\quad+ \lambda^2 \int_{V_0} \left(\widehat{\bs u}^0\cdot \partial_t \bs u_1^{\epsilon}+\widehat{\bs u}^{\epsilon}\cdot \partial_t \bs u_1^{0}  \right)\mathrm{d}V  -\lambda^2 \int_{S_0} f\widehat{\bs u}^0\cdot \partial_t \bs u_1^{0}\,\mathrm{d}S
\ .\quad\quad
\label{Lorentz11}
 \end{eqnarray}
Taking the average in time of \eqref{Lorentz11} over the forcing period, and using that $\bs u_1^{\epsilon}$ and $\bs u_1^0$
are periodic and that
$\bs F_1^{\epsilon}$ and $\bs L_1^{\epsilon}$ are temporal derivatives of  periodic functions  \eqref{NL},  one finally obtains:
 \begin{eqnarray}
& &6\pi\langle\bs U_1^{\epsilon}\rangle_t\cdot\widehat{\bs U}+8\pi\langle\bs \Omega_1^{\epsilon}\rangle_t\cdot\widehat{\bs \Omega}=-{v_1^{\epsilon}}
\label{Re1eps1av1}\ , \quad \mathrm{with}\ ,\\
& &v_1^{\epsilon}={\left\langle \int_{V_0} \Big(\widehat{\bs u}^{\epsilon} \cdot  \bs \nabla \bs u_0^0\cdot \bs u_0^0+\widehat{\bs u}^{0}\cdot[\bs \nabla \bs u_0^{0}\cdot \bs u_0^{\epsilon} +\bs \nabla \bs u_0^{\epsilon}\cdot \bs u_0^{0}]\Big)\mathrm{d}V - \int_{S_0}f \widehat{\bs u}^{0} \cdot  \bs \nabla \bs u_0^0\cdot \bs u_0^0 \mathrm{d}S\right\rangle_t} \quad\quad
\label{Re1eps1av2}
 \end{eqnarray}
  where we have used $\widehat{\bs F}^0=-6\pi\widehat{\bs U}$ and $\widehat{\bs L}^0=-8\pi \widehat{\bs \Omega}$.
Integrating by parts, and using the expressions of $\widehat{\bs u}^0$ and $\bs u_0^0$ (Appendix~\ref{apA}), one obtains:
\begin{eqnarray}
v_1^{\epsilon}=\int_{V_0}\Big(\widehat{\bs u}^{\epsilon}\cdot\bs G_1(\bs r) - {\left\langle\bs u_0^{\epsilon}\cdot \bs G_2(\bs r) \right\rangle_t} \Big)\mathrm{d}V\ ,\quad\quad\label{v1eps}
\end{eqnarray}
with the vector fields $\bs G_1$ and $\bs G_2$ defined as $\bs G_1={\left\langle\bs \nabla{\bs u}_0^0\cdot \bs u_0^0\right\rangle_t}$ and $\bs G_2=[\bs \nabla\widehat{\bs u}^0+(\bs \nabla\widehat{\bs u}^0)^T]\cdot \bs u_0^0$, whose expressions are provided in Appendix~\ref{apA}.  

Using domain perturbation, the velocity field $\bs u_0^{\epsilon}$ (resp. $\widehat{\bs u}^{\epsilon}$) is solution of \eqref{Stokes} (resp. \eqref{unstokes}) with the following boundary conditions on the unit sphere (see Appendix B):
\begin{subeqnarray}
\slabel{u0epsiBC}
\bs u_0^{\epsilon}|_{r=1}&=&-f(\bs n)\partial_r \bs u_0^0|_{r=1}+\bs U_0^{\epsilon} +\bs \Omega_0^\epsilon \times \bs r\\
\widehat{\bs u}^{\epsilon}|_{r=1}&=&-f(\bs n)\partial_r \widehat{\bs u}^0|_{r=1} \ .
\end{subeqnarray} 
 A first simplification comes from recalling  that $\bs \Omega_0^{\epsilon}=\bs 0$. A second one arises from the fact that  the Stokes problem with the uniform boundary condition  $\bs U_0^{\epsilon}$  on the unit sphere does not contribute to particle motion, as demonstrated in Section~\ref{order10}. As a consequence, only the first contribution to $\bs u_0^{\epsilon}|_{r=1}$ in \eqref{u0epsiBC} provides a net contribution to $v_1^\epsilon$.
 
  For clarity, we now distinguish  the cases of pure translation and pure rotation.

\subsubsection{Translation}
Setting $\widehat{\bs \Omega}={\bs 0}$, \eqref{v1eps} simplifies after some algebraic calculations using the definitions of $\bs G_1$, $\bs G_2$, $\widehat{\bs u}^\varepsilon$ and $\bs u_0^\varepsilon$ (Appendices~\ref{apA} and \ref{apB}):
\begin{eqnarray}
v_1^{\epsilon}={\mathcal{K}(\lambda)\left[ f\bs n \bs n \bs n\right]_{\bs n} \vdots\,  \bs{e_xe_x}  \widehat{\bs U}} \ ,
\label{couplingf}
\end{eqnarray}
where $[\bullet]_{\bs n}$ denotes the average over the unit sphere: $[\mathcal{\bullet}]_{\bs n}=\int_{S_0} \mathcal{\bullet}(\bs n) \mathrm dS$, and ${\vdots}$ denotes the three-fold tensorial contraction.    
Quite remarkably,  
 \eqref{couplingf} provides the expression of the net translational velocity as a product of a function of $\lambda$ and a functional of  $f$. The tensorial contraction, together with the angular symmetry properties of the inertial forcing, ensure that only a limited set of the spherical harmonic components of the shape function $f$ can contribute to a net motion.  Further algebraic calculations show that $\mathcal{K}(\lambda)$ conveniently reduces to $\mathcal{K}(\lambda)=\int_{r=1}^{\infty}\frac{\mathrm d\mathcal{J}_{\lambda}(r)}{\mathrm dr}\mathrm{d}r$ with:
\begin{eqnarray}
{\mathcal{J}_{\lambda}(r)}&{=}&{-\frac{1}{4}\Re\Big[\frac{27(1-r^2)}{16\lambda_0^4r^8}\Big(-3|\Lambda_0|^2+2\overline{\Lambda_0}\big(3+3\lambda_0 r-\lambda_0^2 r^2\big)e^{\lambda_0(1-r)}}\nonumber\\
&{\change-}&{\big(3+3\lambda_0r+(\lambda_0r)^2\big)\big(\overline{1+\lambda_0r-\lambda_0^2r^2}\big)e^{2\Re[\lambda_0](1-r)}   \Big)\Big]\ ,}
\end{eqnarray}
where {an overbar denotes the complex conjugate, $\Re[z]$ is the real part operator of $z$} and $\Lambda_0 = 1+ \lambda_0+\lambda_0^2/3$ with $\lambda_0=\lambda e^{-i\pi/4}$. Therefore, using  $\mathcal{J}_{\lambda}(\infty)=\mathcal{J}_{\lambda}(1)=0$, one finds the central result of the present communication:
\begin{eqnarray}
\langle \bs U_1^{\epsilon}\rangle_t=\bs 0\ . \label{zero}
\end{eqnarray}
{No translational net motion can arise at first order (both in Re and non-sphericity $\epsilon$) from geometric asymmetry}. This result stems from the fact that the near-field ($r=O(1)$) and far-field ($r\gg 1$) contributions to the inertial forcing compensate exactly.

\subsubsection{Rotation}
Considering now $\widehat{\bs U}=\bs 0$, the same method  provides:
\begin{eqnarray}
\label{rotation}
v_1^{\epsilon}&=&\mathcal{L}(\lambda)\big[f\bs n\bs n \big]_{\bs n}:{\bs{e_x} (\bs{e_x}\times \widehat{\bs \Omega})}\ ,
\end{eqnarray}
with: {
\begin{eqnarray}
\mathcal{L}(\lambda)=\frac{1}{256} \Im\Big\{\frac1{\Lambda_0}\Big[ \hspace{-0.2cm} && -48 (\lambda_0 (\lambda_0 (\lambda_0 (\lambda_0+6)+18)+30)+24)| \lambda_0| ^2 F(2 \Re(\lambda_0))\nonumber\\
&&+3 i(\lambda_0 (\lambda_0 (\lambda_0 (\lambda_0+9)+27)+42)+30) \lambda_0^2 \bar{\Lambda_0} F(\lambda_0)\nonumber\\
&&+3 i (\lambda_0 (\lambda_0 (\lambda_0(\lambda_0+3)+33)+78)+66) \lambda_0^2 {\Lambda_0}F(i \lambda_0)\nonumber\\
&&+(1-i) \lambda_0^7+(3-7 i) \lambda_0^6-(5+35 i) \lambda_0^5-(6+108 i)
   \lambda_0^4-(60+210 i) \lambda_0^3\nonumber\\
   && -(264+306 i) \lambda_0^2-(348+360 i) \lambda_0-132 i\Big]\Big\} \ ,
   \end{eqnarray}
with $F(z)=[\text{Chi}(z) - \text{Shi}(z)] e^z$ where  $\text{Chi}$/$\text{Shi}$ are the hyperbolic cosine/sine integral functions respectively~\citep{abramowitz1972handbook}. }
We note from \eqref{Re1eps1av2} and \eqref{rotation} that (i) no rotation is obtained along the direction of oscillation (i.e. $\langle\bs \Omega_1^{\epsilon}\rangle_t\cdot\tilde{\bs U}=0$) and that (ii) the particle dynamics is an overdamped rotation toward an equilibrium position. The oscillation direction $\tilde{\bs U}$ is aligned with a principal direction of the symmetric and trace-less second-order tensor $\big[f\bs n\bs n \big]_{\bs n}$ with positive or negative eigenvalue depending on the sign of $\mathcal{L}$.  Further, the function $\langle\mathcal L\rangle_t$ 
changes sign for $\lambda_c \approx 3.6$, resulting in a shift in the equilibrium orientation between $\lambda <\lambda_c$ and $\lambda >\lambda_c$.  This transition confirms fundamental differences in the streaming flow and associated forcing between small and large frequencies, as already observed by \cite{collis2017autonomous} when studying numerically the propulsion of an oscillating asymmetric dumbbell.

\section{Conclusion}\label{ccl}
In this work, we analysed the translation and rotation resulting from the oscillation of a homogeneous near-sphere up to $O(\epsilon\mbox{Re})$, showing analytically that no net translation occurs regardless of the oscillation frequency and despite the geometric asymmetry of the particle. This result, which contradicts the conclusions of \citet{nadal2014asymmetric}, stems from the exact cancellation of the streaming flow forcing in the immediate vicinity of the particle and far away from it, making it difficult to capture numerically as any discretisation introduces necessarily a truncation error.
We also show that a transient rotation can stir back the particle towards one of its equilibrium positions.

Notwithstanding, our results do not contradict  the numerical observations of \citet{collis2017autonomous} for which a weak front-back asymmetry of a dumbbell was sufficient to produce a net motion at that order: in that case, the elongated shape of the particle combined with the small asymmetry of the two spheres leads to an $O(1)$ periodic rotation of the system, which is at the heart of the self-propulsion, when coupled to the oscillating translation -- in contrast, such rotation is absent at $O(\epsilon\mbox{Re})$ in the case of a near-sphere. 
All together, developing net motion around an asymmetric particle appears to require an $O(\epsilon)$ rotation/translation coupling, as obtained for instance using density inhomogeneities.

\section*{Acknowledgements}
We are grateful to F. Nadal and E. Lauga for insightful discussions on this problem. This project has received funding from the European Research Council (ERC) under the European Union's Horizon 2020 research and innovation programme under Grant Agreement 714027 (SM).


\appendix
\section{Unsteady Stokes flow past a spherical particle}
\label{apA}
\subsection{Oscillating flow}
The {complex} velocity field around a sphere oscillating at velocity $\tilde{\bs U}$ reads \citep{kim2013microhydrodynamics} $\bs u_0^0=A\tilde{\bs U}+B(\tilde{\bs U}\cdot \bs n)\bs n$ where:
\begin{eqnarray}
A(r,\lambda)&=&\frac{3}{2\lambda_0^2r^3}\left[-\Lambda_0+\left(1+\lambda_0 r+\lambda_0^2 r^2\right)e^{\lambda_0(1-r)}\right]\ ,\\
B(r,\lambda)&=&\frac{3}{2\lambda_0^2r^3}\left[3\Lambda_0-\left(3+3\lambda_0 r+\lambda_0^2 r^2\right)e^{\lambda_0(1-r)}\right] \ ,
\end{eqnarray}
and where $\lambda_0=\lambda e^{-i\pi/4}$ and $\Lambda_0 = 1+\lambda_0+{\lambda_0^2}/{3}$. Recalling that:
\begin{eqnarray}
\bs \nabla\bs u_0^0&=&A'\tilde{\bs U}\bs n+B'(\tilde{\bs U}\cdot \bs n)\bs{nn}+\frac{B}{r}(\bs I-\bs{nn})(\tilde{\bs U}\cdot \bs n)+\frac{B}{r}\bs n(\bs I-\bs{nn})\cdot \tilde{\bs U}\ ,
\end{eqnarray}
one may compute $\bs G_1={\left\langle\bs \nabla\bs u_0^0\cdot \bs u_0^0\right\rangle_t}$, that is:
\begin{eqnarray}
{\bs G_1}&=&{\frac{1}{2}\Re\Big[(A+B)\overline{[A'\bs I\bs n +B'\bs n\bs n\bs n]} +\frac{A\overline{B}}{r}[\bs n(\bs I-\bs n\bs n)+(\bs I-\bs n\bs n)\bs n]\Big]:\bs{e_xe_x}}\ .
\end{eqnarray}
\subsection{Steady Translation}
The particular case of a steady translating sphere ($\lambda=0$) at velocity $\widehat{\bs U}$ is given by $\widehat{\bs u}^0=\widehat{A}\widehat{\bs U}+\widehat{B}(\widehat{\bs U}\cdot \bs n)\bs n$ where:
\begin{eqnarray}
\widehat{A}(r)&=&\frac{3}{4r}+\frac{1}{4r^3}\ ,\quad\quad \widehat{B}(r)=\frac{3}{4}\Big(\frac{1}{r}-\frac{1}{r^3}\Big)\ .
\end{eqnarray}
One may compute
 $\bs G_2=\bs \nabla\widehat{\bs u}_0^0\cdot \bs u_0^0+\bs u_0^0\cdot \bs \nabla\widehat{\bs u}_0^0$, that is:
\begin{eqnarray}
\bs G_2&=&\widehat{A}'A\bs n\bs I: \widehat{\bs U}\tilde{\bs U} + \widehat{A}'(A+B)\bs I\bs n:\tilde{\bs U}\widehat{\bs U} + (\widehat{A}'B+2\widehat{B}'(A+B))\bs n \bs n\bs n: \tilde{\bs U}\widehat{\bs U}\nonumber\\
&+& \frac{\widehat{B}A}{r}\Big[\bs n(\bs I-\bs n\bs n):\tilde{\bs U}\widehat{\bs U}+2(\bs I-\bs n \bs n)\bs n:\widehat{\bs U}\tilde{\bs U}\Big]+\frac{\widehat{B}(A+B)}{r}(\bs I-\bs n\bs n)\bs n: \tilde{\bs U}\widehat{\bs U}\ . \quad\quad 
\end{eqnarray}

\subsection{Steady Rotation}
The velocity field around a steady rotating sphere reads $\widehat{\bs u}^{0}={\widehat{\bs \Omega}\times\bs n}/{r^2}$.
Computing: 
\begin{eqnarray}
\bs \nabla\widehat{\bs u}^{0}\cdot \bs u_0^{0}&=&\frac{1}{r^3}[\widehat{\bs \Omega}\times\bs u_0^{0}-3(\bs u_0^{0}\cdot \bs n)(\widehat{\bs \Omega}\times\bs{n})]\ ,\\
\bs u_0^{0}\cdot\bs \nabla\widehat{\bs u}^{0}&=&\frac{1}{r^3}[\bs u_0^{0}\times\widehat{\bs \Omega}-3\bs u_0^{0}\cdot(\widehat{\bs \Omega}\times\bs{n})\bs n] \ ,
\end{eqnarray}
one obtains the expression of $\bs G_2=\bs \nabla\widehat{\bs u}_0^0\cdot \bs u_0^0+\bs u_0^0\cdot \bs \nabla\widehat{\bs u}_0^0$ as:
\begin{eqnarray}
\bs G_2&=& -\frac{3A}{r^3}(\tilde{\bs U}\times\widehat{\bs \Omega})\cdot \bs {nn}- \frac{3(A+B)}{r^3}(\tilde{\bs U}\cdot \bs n)(\widehat{\bs \Omega}\times \bs n)\ .
\end{eqnarray}

\section{Unsteady Stokes flow past a nearly-spherical particle}
\label{apB}
Here we compute the velocity field solution of the unsteady Stokes problem around a nearly-spherical particle:
\begin{align}
\lambda^2\partial_t \bs u_0^{\epsilon}=-\bs \nabla p_0^{\epsilon}+\bs \nabla^2\bs u_0^{\epsilon}\ ,&\qquad 
\bs\nabla\cdot\bs u_0^{\epsilon}=0 \ ,\\
\label{bc0}
\bs u_0^{\epsilon}|_{r=1}=-f(\bs n)\partial_r\bs u_0^0|_{r=1}\ ,&\qquad 
\bs u_0|_{\bs r \rightarrow \infty}= \bs 0 \ .
\end{align}
{In  Fourier space} the boundary condition on the surface of the particle \eqref{bc0} takes the form \citep{zhang1998oscillatory}:
\begin{eqnarray}
\bs u_0^{\epsilon}|_{r=1}&=&\frac{3f(\bs n)}{2}(1+\lambda_0)(\bs I-\bs{nn})\cdot {\tilde{\bs U}} \ .
\end{eqnarray}
Following \citet{sani1963convective}, we perform a reconstruction of the velocity field from its radial component and associated vorticity: 
\begin{eqnarray}
\bs u_0^{\epsilon}=u_{0,r}^{\epsilon}\bs n+r^2\sum_{n=1}^{\infty}\frac{1}{n(n+1)}\left[ \bs\nabla_s(\nabla^2u_{r,n})-\bs n\times\bs\nabla_s\chi_{r,n}\right] \ ,
\end{eqnarray}
where $\bs \nabla_s = \bs \nabla - \bs n\partial_r$ and where $u_{r,n}$ denotes the $n^\mathrm{th}$ mode in the spherical harmonics basis of the radial component of $\bs u_0^{\epsilon}$.
The latter satisfies in time-Fourier space the equation $\bs\nabla^2(i\lambda^2+\bs\nabla^2)(r u_{0,r}^{\epsilon})=0$. The function $\chi_{r,n}$ is the $n^\mathrm{th}$ mode of the radial component of $\bs \nabla \times \bs u_0^{\epsilon}$, satisfying $(\bs\nabla^2+i\lambda^2)(r\chi^{\epsilon}_r)=0$.
 Defining $p$ and $q$ through:
\begin{equation}
 p=-\frac{2\bs\nabla_s\cdot\bs u_0^{\epsilon}|_{r=1}}{3(1+\lambda_0)}\ , \quad \quad
q=\frac{2\bs n\cdot\bs\nabla_s\times \bs u_0^{\epsilon}|_{r=1}}{3(1+\lambda_0)} \ ,\label{pq}
\end{equation}
one finally obtains: 
\begin{equation}
\bs u_0^{\epsilon}=\frac{1}{r}\sum_{n=1}^{\infty}\sum_{m=-n}^{n}U_{n}p_{n}^mY_{n}^m\bs n+\sum_{n=1}^{\infty}\sum_{m=-n}^{n}\frac{r^2V_{n}p_{n}^m}{n(n+1)}\bs\nabla Y_{n}^m- \sum_{n=1}^{\infty}\sum_{m=-n}^{n}\frac{rX_{n}q_{n}^m}{n(n+1)}\bs n\times\bs \nabla Y_{n}^m \ ,\quad 
\label{generalLamb}
\end{equation}
where $p_{n}^m$ and $q_{n}^m$ denote respectively the modes of $p$ and $q$ in the spherical harmonics basis $(Y_{n}^m)$, and $U_n$, $V_n$ and $X_n$ follow:
\begin{eqnarray}
U_n(r,\lambda)&=&\frac{3}{2}(1+\lambda_0)\frac{h_{n}^{(1)}(\overline{\lambda_0}r)-\frac{h_{n}^{(1)}(\overline{\lambda_0})}{r^{n+1}}}{(2n+1)h_{n}^{(1)}(\overline{\lambda_0})-\overline{\lambda_0}h_{n+1}^{(1)}(\overline{\lambda_0})}\ ,\\
V_n(r,\lambda)&=&\frac{U_n(r,\lambda)}{r^2}+\frac{\partial_rU_n(r,\lambda)}{r}\ ,\\%
X_n(r,\lambda)&=&\frac{3}{2}(1+\lambda_0)\frac{h_{n}^{(1)}(\overline{\lambda_0}r)}{h_{n}^{(1)}(\overline{\lambda_0})}\ ,
\end{eqnarray}
with $h_{n}$  the spherical Hankel function of the first kind and order $n$ \citep{abramowitz1972handbook}. In \eqref{pq}, the functions $p$ and $q$ defined on the surface of the unit sphere are directly related to the shape function $f$ through \eqref{bc0}. Using \eqref{volume}, they further satisfy : 
\begin{align}
[p]_{\bs n}&=[q]_{\bs n}=0\ ,\qquad [q \bs n]_{\bs n}=0\ ,\\
[\nabla p \bs n]_{\bs n}&=[\bs n \nabla p]_{\bs n}=[p \bs n\bs n]_{\bs n}=-2[f \bs n\bs n\bs n]_{\bs n}{\cdot}\tilde{\bs U}\ .
\end{align}
Note that these results can be transposed to obtain $\widehat{\bs u}^{\epsilon}$ taking $\lambda=0$ for the translation problem. And a similar approach can be used in the rotating case.

\bibliographystyle{jfm}
\bibliography{biblio}

\end{document}